\documentclass[longauth]{aa} 

\newcommand{\degree}{\ensuremath{^\circ}}                                   
\usepackage{natbib}
\usepackage{rotating}
\usepackage{graphicx}
\usepackage{lscape}
\usepackage{txfonts}

\usepackage{txfonts}
\begin{document}
   \title{MAGIC long-term study of the distant TeV blazar PKS~1424+240 in a multiwavelength context}   


\author{
J.~Aleksi\'c\inst{1} \and
S.~Ansoldi\inst{2} \and
L.~A.~Antonelli\inst{3} \and
P.~Antoranz\inst{4} \and
A.~Babic\inst{5} \and
P.~Bangale\inst{6} \and
U.~Barres de Almeida\inst{6} \and
J.~A.~Barrio\inst{7} \and
J.~Becerra Gonz\'alez\inst{8,}\inst{25,*} \and
W.~Bednarek\inst{9}
% ...
}
\institute { IFAE, Campus UAB, E-08193 Bellaterra, Spain
\and Universit\`a di Udine, and INFN Trieste, I-33100 Udine, Italy
\and INAF National Institute for Astrophysics, I-00136 Rome, Italy
\and Universit\`a  di Siena, and INFN Pisa, I-53100 Siena, Italy
\and Croatian MAGIC Consortium, Rudjer Boskovic Institute, University of Rijeka and University of Split, HR-10000 Zagreb, Croatia
\and Max-Planck-Institut f\"ur Physik, D-80805 M\"unchen, Germany
\and Universidad Complutense, E-28040 Madrid, Spain
\and Inst. de Astrof\'isica de Canarias, E-38200 La Laguna, Tenerife, Spain
\and University of \L\'od\'z, PL-90236 Lodz, Poland
\and Deutsches Elektronen-Synchrotron (DESY), D-15738 Zeuthen, Germany
\and ETH Zurich, CH-8093 Zurich, Switzerland
\and Universit\"at W\"urzburg, D-97074 W\"urzburg, Germany
\and Centro de Investigaciones Energ\'eticas, Medioambientales y Tecnol\'ogicas, E-28040 Madrid, Spain
\and Technische Universit\"at Dortmund, D-44221 Dortmund, Germany
\and Inst. de Astrof\'isica de Andaluc\'ia (CSIC), E-18080 Granada, Spain
\and Universit\`a di Padova and INFN, I-35131 Padova, Italy
\and Universit\`a dell'Insubria, Como, I-22100 Como, Italy
\and Unitat de F\'isica de les Radiacions, Departament de F\'isica, and CERES-IEEC, Universitat Aut\`onoma de Barcelona, E-08193 Bellaterra, Spain
\and Institut de Ci\`encies de l'Espai (IEEC-CSIC), E-08193 Bellaterra, Spain
\and Japanese MAGIC Consortium, Division of Physics and Astronomy, Kyoto University, Japan
\and Finnish MAGIC Consortium, Tuorla Observatory, University of Turku and Department of Physics, University of Oulu, Finland
\and Inst. for Nucl. Research and Nucl. Energy, BG-1784 Sofia, Bulgaria
\and Universitat de Barcelona, ICC, IEEC-UB, E-08028 Barcelona, Spain
\and Universit\`a di Pisa, and INFN Pisa, I-56126 Pisa, Italy
\and now at: NASA Goddard Space Flight Center, Greenbelt, MD 20771, USA and Department of Physics and Department of Astronomy, University of Maryland, College Park, MD 20742, USA
% ...
}
  \date{Received January 1, 2014; accepted April 29, 2014}

  \abstract
    {}
   {We present a study of the very high energy (VHE; E$>$100~GeV) $\gamma$-ray emission of the blazar PKS\,1424+240 observed with the MAGIC telescopes. The primary aim of this paper  is the multiwavelength spectral characterization and modeling of this blazar, which is made particularly interesting by the recent discovery of a lower limit of its redshift of z\,$\geq$\,0.6 and makes it a promising candidate to be the most distant VHE source.}
   {The source has been observed with the MAGIC telescopes in VHE $\gamma$ rays for a total observation time of $\sim$33.6\,h from 2009 to 2011. A detailed analysis of its $\gamma$-ray spectrum and time evolution has been carried out. Moreover, we have collected and analyzed simultaneous and quasi-simultaneous multiwavelength data.}
   {The source was marginally detected in VHE $\gamma$ rays during 2009 and 2010, and later, the detection was confirmed during an optical outburst in 2011. The combined significance of the stacked sample is $\sim$7.2\,$\sigma$. The differential spectra measured during the different campaigns can be described by steep power laws with the indices ranging from 3.5~$\pm$~1.2 to 5.0~$\pm$~1.7. 
 The  MAGIC spectra corrected for the absorption due to the extragalactic background light connect smoothly,  within systematic errors, with the mean spectrum in 2009-2011 observed at lower energies by the {\it Fermi}-LAT. The absorption-corrected MAGIC spectrum is flat with no apparent turn down up to 400\,GeV. 
The multiwavelength light curve shows increasing flux in radio and optical bands that could point to a common origin from the same region of the jet. The large separation between the two peaks of the  constructed non-simultaneous spectral energy distribution also requires an extremely high Doppler factor if an one zone synchrotron self-Compton model is applied. We find that a two-component synchrotron self-Compton model describes the spectral energy distribution of the source well, if the source is located at $z\sim0.6$.} 
   {} 
   \titlerunning{MAGIC long-term study of the distant TeV blazar PKS~1424+240}
   \authorrunning{Aleksi\'c et al.}
   \maketitle
%

\section{Introduction}
Blazars are active galactic nuclei (AGN) that host a relativistic jet,
which is pointed   at a small angle to the line of sight. The
spectral energy distribution (SED) of blazars shows a two-bump
structure. It is widely accepted that the lower energy bump is due to
synchrotron emission produced by the relativistic electrons spiraling
in the magnetic field of the jet. The location of the peak of this
lower energy bump in the SED is used to classify the sources 
as low, intermediate, and high-synchrotron-peaked blazars
\citep{Fermi_SED}.

The high energy-peaking blazars are the most numerous, extragalactic
very high-energy (VHE, E $>$ 100\,GeV) $\gamma$-ray sources. The
origin of the VHE $\gamma$-ray emission is still under debate. It is
typically modeled with  synchrotron self-Compton (SSC) emission
models where the synchrotron radiation serves as seed photons for
inverse Compton scattering \citep{maraschi}. 
However, hadronic processes, such as proton synchrotron and radiation produced by the secondary 
particles, can also produce the observed VHE $\gamma$-ray
emission  \citep[e.g.,][]{mannheim92,aharonian00}.

Both models fit the observed IR, optical, X-ray, and $\gamma$-ray data well 
\citep[e.g.,][]{tavecchio,reimer}, while it is generally assumed that 
the emission region is still opaque in radio bands and that the radio 
emission originates from a different emission region \citep{katarzynski01}. 

The object \object{PKS 1424+240} was discovered in the  1970s as a radio source \citep{fanti94} 
and was identified as a blazar by \citet{impey}.
The source was detected in $\gamma$ rays by  the {\it Fermi} Large Area Telescope \citep[LAT; ][]{atwood09} with a very hard spectrum with a photon index of $\Gamma=1.85\pm0.07$ 
\citep{Fermi_spectrum}, and 
it entered the family of VHE $\gamma$-ray emitters in spring 2009 when a
 detection was first reported by VERITAS \citep{ver_atel} 
and soon after confirmed by MAGIC \citep{magic_atel}. 
The source was previously observed during 2006 and
2007 by the MAGIC-I telescope, and a flux upper limit of 8.2 \% Crab Units\footnote{The Crab unit used in this work is an arbitrary unit obtained by dividing the integral energy flux measured above a certain threshold by the Crab Nebula flux, which was measured above the same threshold by MAGIC \citep{analysis1}.} for
E$>120$ GeV was derived \citep{stacking}. 
 The VERITAS observations in 2009 indicated  a  steady flux with the photon spectrum, which is well described by a power law with a photon index of $3.8 \pm 0.5_{\mathrm stat} \pm 0.3_{\mathrm syst}$ and a flux normalization at 200\,GeV of  $(5.1 \pm 0.9_{\mathrm stat}\pm 0.5_{\mathrm syst}) \cdot 10^{-11}$ TeV$^{-1}$ cm$^{-2}$ s$^{-1}$ \citep{acciari10}.
 As this manuscript was being resubmitted to the journal, the VERITAS collaboration reported observations of PKS1424+240 from 2009, 2011 and 2013, confirming the soft spectrum that had been reported with the 2009 dataset but indicated significant gamma-ray flux variability \citep{archambault14} this time.

 The source \object{PKS 1424+240} is a BL Lac object, which by definition shows weak or no emission lines in its optical  spectra. 
Therefore, like for many BL Lacs, the redshift of \object{PKS 1424+240}
is still uncertain. \citet{rau12} reported a photometric upper limit of z$_{ul}$~=~1.1.
Recently \citet{furniss} determined a lower limit of the redshift z $\geq$ 0.6035 from the Ly$\beta$ and Ly$\gamma$ absorption.
This is more distant than 3C~279 (z~=~0.536), which was long considered to be the most distant VHE $\gamma$-ray emitter. 
   Even though there are other sources with high lower limits on the redshift\footnote{e.g. KUV~00311-1938, z$>0.506$  \citep{pita12}}, this makes \object{PKS 1424+240} a strong candidate to be the most distant known VHE gamma-ray emitter.
Later, we discuss the  redshift constraints obtained from VHE $\gamma$-ray data analysis, which seem to confirm this result. 

It should be noted that \citet{meisner10} report a detection of the host galaxy in the i-band,
 and assuming that the galaxies hosting BL Lac
objects can be considered as standard candles \citep{sbarufatti05}, they estimated the redshift to be z$=0.23^{+0.06}_{-0.05}$.
However, this value is in conflict with a new photometric lower limit from Shaw et al. (2013) and with the spectroscopic limit reported in Furniss et al. (2013).

In this paper, we present MAGIC observations of the source that  include
the first detection of the source in 2009, follow-up observations in
2010, and target of opportunity observations triggered by the optical
high state of the source in Spring 2011. The
differential and integral energy spectra are presented  with a
study of the spectral variability. 
We carried out an extensive multiwavelength study which makes use of the data available from $\gamma$ rays to radio  to study multiwavelength properties and model the SED of the source.

\section{MAGIC observations and results}
\label{sec:observations}

\subsection{MAGIC data}
MAGIC (Major Atmospheric Gamma-ray Imaging Cherenkov) 
is a system of two 17~m Imaging 
Atmospheric Cherenkov Telescopes (IACT) located at the Roque de los Muchachos, 
Canary Island of La Palma at the height of 2200~m above sea level. 
The commissioning of 
the second MAGIC telescope finished at the end of 2009, and since then, both
telescopes have worked together in stereoscopic mode \citep{cortina,performance_stereo}.  
In simple terms, MAGIC observes the faint Cherenkov light emitted in the
atmosphere by a shower of particles that is induced by a VHE $\gamma$ ray when
entering the atmosphere.
The light is focused into a camera and forms an image, 
which is triggered, registered, calibrated, and then 
parametrized with the so-called {\it Hillas parameters} \citep{hillas}.
These parameters are used to separate the $\gamma$-like
events, which constitute the signal, from the background 
dominated by hadronic events. 
In addition, dedicated Monte Carlo simulations of the system 
performance are used for the energy reconstruction and the 
$\gamma$-hadron separation. 

The source \object{PKS 1424+240} was observed
in single-telescope mode (i.e. using only MAGIC-I) from April 
to June 2009, when MAGIC~II was in commissioning phase, and observed in stereo 
mode in Spring 2010 and Spring 2011. 
Each data sample was analyzed independently since the 
performance of the instrument changed over the years. 

The source \object{PKS 1424+240} was observed in good  conditions for 12.5\,h in 2009 (from MJD~54938 to MJD~55005), 
covering a zenith angle range between 4\degree~and 36\degree. 
In addition, stereo data were collected at the same zenith angle range in early 2010
 between March to April (from MJD~55269 to MJD~55305)
for a total observation time of 
$\sim$11.6\,hours and 9.5\,hours from April to May 2011 (from MJD~55875 to MJD~55889). 
The whole data sample was taken in the false-source tracking (wobble) mode 
\citep{wobble}, in which the telescopes were alternated every 20\,minutes 
between two symmetric sky positions at $0.4^\circ$ offset from the source. 
The wobble enables us to take both source and background data simultaneously.

The analysis of the data was performed using the standard MAGIC analysis and reconstruction software 
\citep{analysis1,analysis2, analysis3, analysis4}. 
In the analysis of stereo data presented in this work, we took advantage of
a direction reconstruction method based on the
{\it DISP RF} method, as described in  \citet{aleksic10}, and
adapted to the stereo observations (see  \cite{performance_stereo} for more details). 

\subsection{Signal search}
The signal search is performed by making use of the distribution of the 
$\theta^2$ parameter, which is 
defined as the square of the angular distance between the
reconstructed shower direction in the telescope camera
and the real position of the source. 
To avoid systematic effects
in the background calculation, we considered  only events with 
the parameter {\it size}\footnote{The parameter {\it size} represents the total number of photo-electrons in an image.} 
that is larger than 200\,photo-electrons (mono analysis) 
and 50\,photo-electrons (stereo analysis). 
The corresponding energy thresholds 
(defined as the energies where the number of Monte Carlo $\gamma$-ray events with assumed spectral 
indices is maximized in the histogram of the image {\it size}) 
are $\sim$150\,GeV for 2009 mono observations and
$\sim$100\,GeV for 2010 and 2011 stereoscopic observations.

The signal is extracted by comparing the $\theta^2$
distribution of the source region ({\it ON}) 
with the background ({\it OFF}), which is estimated from one 
(in 2010 and 2011 data samples) and three regions (in the 2009 data sample) 
of the sky located near the source of interest and collected 
simultaneously with the source observation.
The significance of the detection for the different data samples can be found in Table~\ref{table_significance}.
During the 2009 observational campaign, a hint of a signal was found with a number of 
excess events  $N_{ex}\approx498$  ($N_{ex}=N_{ON}-N_{OFF}$) and 
4.6\,$\sigma$, which are calculated according to 
eq.\,17 of \citet{lima83}. This hint of signal was confirmed later in the 2010 and 2011 campaigns: 
the $\theta^2$ distribution from the 2010 sample gives $\sim$330 excess events with a 
significance of 4.8\,$\sigma$, while  
an excess of  $N_{ex}$$\sim$333 was found corresponding to a significance of  5.5\,$\sigma$ in the 2011 sample.
The stacked significance of the overall sample (33.6 hours of data from 2009 to 2011, 1161 excess events) is more than 7\,$\sigma$. 

\begin{table}[b]
\begin{center}
\caption{PKS 1424+240 observation characteristics and signal search results. The year of observation, the observation time of the final selected sample, the energy threshold (E$_{th}$), the number of excess events (N$_{exc}$), and the significance of the signal are reported.\label{table_significance}}
\begin{tabular}{cccccc}
\hline
\hline
Year   & Obs. Time &E$_{th}$  & N$_{exc}$ & Significance\\
\hline
\hline
2009   & 12.5 h   & $\sim$150\,GeV &  498 & 4.6\,$\sigma$\\
2010   & 11.6 h   & $\sim$100\,GeV &  330 & 4.8\,$\sigma$\\
2011   & 9.5 h   & $\sim$100\,GeV &  333 & 5.5\,$\sigma$\\
\hline
\hline
\end{tabular}
\end{center}
\end{table}

\begin{table*}[ht]
\begin{center}
\caption{PKS 1424+240 spectrum power-law fit parameters and integral flux values.\label{tab:spectrum_fits}}
\begin{tabular}{ccccc}
\hline
\hline
Period & Fit range & $f_0$                                   & $\Gamma$& $F$($>$150\,GeV)                         \\ 
           
       &     [GeV] & [cm$^{-2}$ s$^{-1}$ TeV$^{-1}$]    &  & [cm$^{-2}$ s$^{-1}$]  \\  
\hline
\hline
2009   & 150 -- 400 &(1.3 $\pm$ 0.6$_{stat}$ $\pm$ 0.4$_{sys}$) $\cdot$ 10$^{-10}$  &  5.0 $\pm$ 1.7$_{stat}$ $\pm$ 0.7$_{sys}$& (1.66 $\pm$ 0.50) $\cdot$ 10$^{-11}$\\
2010   & 100 -- 300  &(0.5 $\pm$ 0.2$_{stat}$ $\pm$ 0.1$_{sys}$) $\cdot$ 10$^{-10}$  &  3.5 $\pm$ 1.2$_{stat}$ $\pm$ 0.5$_{sys}$&(0.53 $\pm$ 0.25) $\cdot$ 10$^{-11}$ \\
2011   & 100 -- 400 &(1.0  $\pm$ 0.3$_{stat}$ $\pm$ 0.2$_{sys}$) $\cdot$ 10$^{-10}$ &  3.9 $\pm$ 0.7$_{stat}$ $\pm$ 0.2$_{sys}$ &(1.00 $\pm$ 0.30) $\cdot$ 10$^{-11}$\\
\hline
\hline
\end{tabular} 
\end{center}
\end{table*}

\subsection{Differential energy spectra}

\begin{figure}[b!]
 \centering
  \includegraphics[angle=0,width=0.47\textwidth]{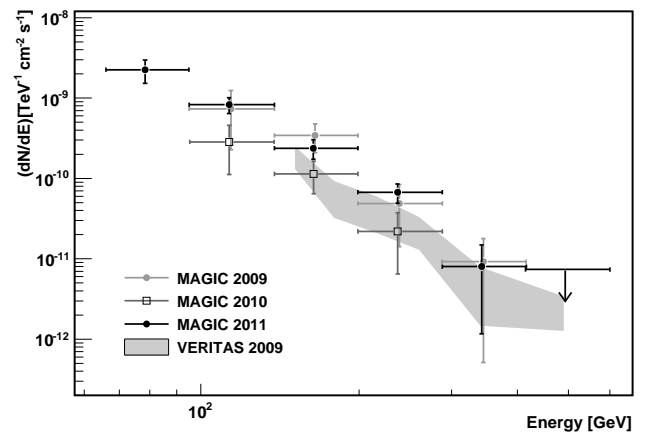}
  \caption{Differential energy spectra of PKS 1424+240 measured by MAGIC in 2009 (gray circles), 2010 (squares), and 2011 (black circles). The black downward arrow represents the 95\% confidence level upper limit for 2011 data.  The gray-shaded area represents the results obtained in 2009 with VERITAS \citep{acciari10}. \label{spectrum}}
\end{figure}

The differential energy spectra observed with the MAGIC telescopes in 2009, 2010, and 2011 campaigns are shown in Figure~\ref{spectrum}.
In each case, the spectrum can be well 
fit with a simple power law of the form:
\begin{equation}
\frac{dN}{dE}=f_0 \cdot \left(\frac{E}{\mathrm{200 \, \mathrm{GeV}}} \right)^{-\Gamma} \hspace{0.5cm}  [\frac{\mathrm{ph}}{\mathrm{cm}^2 \cdot \mathrm{s} \cdot \mathrm{TeV}}] .
\end{equation}

The parameters resulting from the fits are compatible within the uncertainties and are reported in Table~\ref{tab:spectrum_fits}. The systematic uncertainties in the spectral measurements with MAGIC for spectra similar to that of the Crab Nebula (i.e. bright and relatively hard spectra) were reported in \cite{MAGICcrab} and \cite{performance_stereo} for mono and stereo observations, which were about 11\% in the normalization factor (at 300 GeV) and 0.15--0.20 in the photon index. 
However, for spectral shapes described by a photon index of 4 and gamma-ray fluxes lower than that of the Crab by one order of magnitude, the systematic errors increase substantially. Considering the spectral shape of \object{PKS 1424+240}, the relatively low dynamic range over which MAGIC detects photons (100--400 GeV), and the relatively low signal-to-noise background of the different datasets, we estimated the systematic uncertainties to be 30\% in the flux level and 0.7 in the power-law index for the 2009 dataset (obtained with mono observations), 20\% and 0.54 for the flux and the power-law index for the 2010 dataset (obtained with stereo observations) respectively, and 19\% and 0.22 for the flux and the power-law index for the 2011 dataset (obtained with stereo observations, where the gamma-ray flux is about twice that measured in 2010), respectively.  The error on the flux does not include uncertainty on the energy scale.  The energy scale of the MAGIC telescopes is determined with a precision of about 17\% at low energies (E \textless 100 GeV) and 15\% at medium energies (E \textgreater 300 GeV) \citep{performance_stereo}. 


The mean differential flux registered by VERITAS between February to June 2009
is significantly lower from the spectrum observed by MAGIC~I in the same period. 
This apparent discrepancy is likely related 
to the different time coverage of the two observations
and is discussed later. 
The spectral slope we derive is consistent with the results presented in \cite{acciari10}. 
However, unlike the method in \cite{acciari10}, we do not detect excess events above 
400\,GeV in any data sample. 
We therefore derive a 95\% confidence level upper 
limit for the energy bin $416-601$\,GeV 
of $7.2\cdot10^{-12}$ TeV$^{-1}$cm$^{-2}$s$^{-1}$ for 2011 data. 
This limit agrees with the 
value measured by \citet{acciari10} at $\sim$\,500 GeV.

\subsection{VHE variability analysis}

\begin{figure}[h]
\centering 
\includegraphics[width=0.48\textwidth]{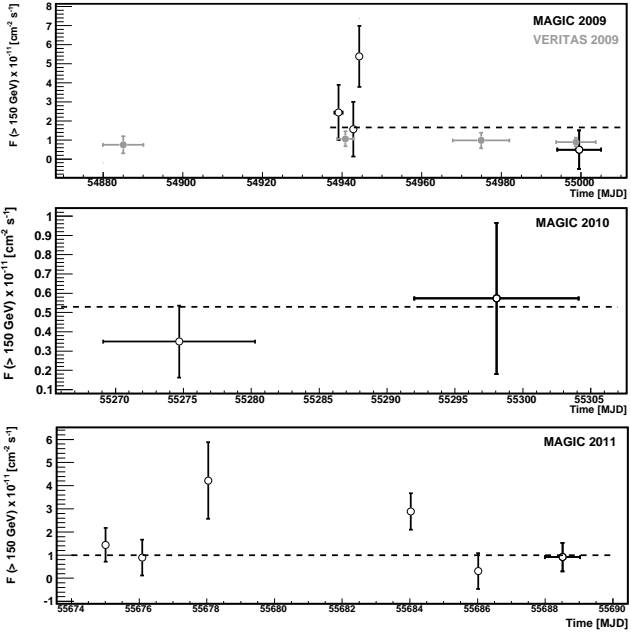}
\caption{VHE $\gamma$-ray light curve of PKS 1424+240 for E$>$150\,GeV from the observation campaigns in 2009, 2010 and 2011 with the MAGIC telescopes. Dashed lines represent  the mean MAGIC fluxes, while  the gray points represent the light curve measured by VERITAS in 2009, as adapted from \citet{acciari10}}. \label{fig:TeV_lcs}
\end{figure}

Table~\ref{tab:spectrum_fits} reports the integral fluxes  measured by MAGIC from 2009 to 2011 above 150\,GeV, which are 
drawn in Fig.~\ref{fig:TeV_lcs} (dashed lines). 
 All the fluxes are below the 
upper limit derived from the previous MAGIC observation in 2006 and 2007, which was (above 120\,GeV at 95\% confidence level) $3.1\cdot10^{-11}$cm$^{-2}$s$^{-1}$ \citep{stacking}.
 
A constant flux hypothesis is consistent with the yearly values from 2009 to 2011 at 95\% confidence level  ($\chi^2/d.o.f.=4.4/2$). 
In addition, constant fits to the light curves of the
individual years also agree with the hypothesis
of constant flux at the same confidence level. 

 Interestingly, VERITAS found no variability
in the 2009 dataset in the energy range of 140\,GeV--600\,GeV. 
In Figure \ref{fig:TeV_lcs}, the gray markers refer to VERITAS 2009 observations from \citet{acciari10}. Since VERITAS reported the integral flux in the range 140\,GeV--600\,GeV, the fluxes have been scaled down by 18\%  (derived from the spectral index of $\Gamma$=\,3.8 measured by VERITAS) to be compared to the MAGIC light curve at above 150 GeV.
The statistical errors of the MAGIC flux measurements are larger than those from VERITAS because the MAGIC 2009 observations were performed with a single telescope\footnote{
The MAGIC Stereo sensitivity is a factor of 3 better than the one of the MAGIC I telescope at the lowest VHE energies \citep{performance_stereo}.} and were substantially shorter (12 hours vs 28 hours) than those from VERITAS.


\subsection{Redshift estimation from $\gamma$-ray observations}

The redshift of \object{PKS 1424+240} is still uncertain and can  be constrained
by VHE $\gamma$-ray observations by considering
the interaction of energetic photons with the 
diffuse optical and near-infrared background, which are also called the
extragalactic background light (EBL) \citep{hauser01}.
The interaction of VHE photons with the EBL leads to electron-positron 
pair production. 
The effect is a partial or total absorption of the VHE photons coming 
from distant sources
and a consequent distortion of the emitted spectrum.
Due to the large uncertainties in the direct EBL measurements, 
many models have been proposed in recent years 
\citep{stecker06,franceschini08,gilmore09,kneiske10,dominguez, stecker12}.

In \citet{acciari10}, the authors present the photon index 
measured with {\it Fermi}-LAT at low energies
in combination with VERITAS data 
corrected by the EBL absorption effect using three recent EBL models 
\citep{franceschini08,gilmore09,finke10} and 
derive an upper limit of the redshift of z$<$0.66 with a 95\% confidence level.

A similar approach is undertaken in \citet{prandini10,prandini11},
where the idea that the slope of the VHE spectrum corrected for EBL absorption 
should not be harder than the one measured by {\it Fermi}-LAT at lower energies is tested 
on a sample of known redshift sources.
The redshifts $z^*$ at which the two slopes match 
are compared to the spectroscopic redshifts, $z_{spec}$, 
with the result that $z^*$ is above $z_{spec}$ in all the cases considered. 
Therefore, $z^{*}$ can be used as upper limit estimate of the source distance 
if there is no spectroscopic redshift available.

If we apply the same method to the data  presented here, and 
in particular, to the 2011 spectral points that are the most precise,
we obtain $z^*$=0.61~$\pm$~0.10, where $z^*$ is the redshift at which
the EBL--deabsorbed spectrum has the same slope as measured
by {\it Fermi}-LAT at lower energies, by assuming the \citet{franceschini08} EBL model. 
This implies a 2$\sigma$ upper limit on the redshift of 0.81. 

The upper limit derived agrees with the  value obtained by \cite{furniss}, and, 
therefore, we hereafter adopt the redshift of $z=0.6$ for the source
 to not overestimate the EBL absorption.

\section{Multiwavelength view of  PKS 1424+240}

The multiwavelength data for \object{PKS 1424+240} were collected from several
ongoing monitoring programs in radio, optical, X-rays and high-energy
$\gamma$ rays. The resulting multiwavelength light curves are shown in
Figure~\ref{fig:lc}.

\subsection{Radio data}
The object \object{PKS 1424+240} is monitored at 15\,GHz using the 40 meter telescope of
the Owens Valley Radio Observatory as a part of a larger monitoring
program, where a sample of $\sim$ 1700 sources is observed twice a
week 
\citep{richards11}. The telescope is equipped with
dual-beamed, off-axis optics and a cooled receiver installed at the
prime focus. The two sky beams are Dicke switched using the off-source
beam as a reference, and the source is alternated between the two
beams in an ON-ON fashion to remove atmospheric and ground
contamination. The calibration reference source is 3C~286 for which the flux
density of 3.44\,Jy at 15\,GHz is assumed \citep{baars77}. 
The
systematic uncertainty of about 5\% in the flux density scale is not
included in the error bars. Details of the observations, calibration, and
analysis are given in \cite{richards11}.

Visual inspection of the 15\,GHz light curve shows an increasing trend
from 2009 to 2011; the average flux increases 
from  0.26$\pm$0.01\,Jy (MJD\,54923) to 0.31$\pm$0.01\,Jy (MJD\,55807)
with minimum and maximum values of 0.20$\pm$0.01\,Jy (MJD\,54954) and 0.32$\pm$0.02\,Jy (MJD\,55721), respectively.
The constant hypothesis can be discarded due to its low probability 
($\chi^2/d.o.f.=861/126 $, probability smaller than 0.0001).

We test the trend by using the five methods for obtaining the
linear regression fits from \cite{isobe}
and find the trend to be significant at the $>20\sigma$ level. 
We also calculate the intrinsic modulation index of the light curve using the likelihood method introduced in \cite{richards11}. The intrinsic modulation index is a measure of variability, similar to the standard variability index  \citep[e.g.][]{aller92} but takes the errors and sampling into account in the calculation of the likelihood {\citep[see][for further details]{richards11}}. For the radio data, we obtain a modulation index of $0.072\pm0.005$. This value confirms that the source is significantly variable and the variability amplitude is close to the mean value of 0.036$\pm$0.010 obtained for a sample of 98 HSP BL Lacs in the OVRO sample \citep{richards14}.

The visual inspection also suggests a 
small amplitude flare in 2011. The existence of this flare was tested
by removing the increasing trend by fitting a line to data and then
calculating the intrinsic modulation index using the likelihood method. Because this method takes the flux density of the source into account, the average flux density 0.28\,Jy was added to the residuals. The intrinsic modulation index for the de-trended data is then $0.026\pm0.003$, implying that the
light curve is indeed variable above the 3$\sigma$ level even if the
trend is removed. 

The source has also been observed by the Mets\"ahovi 13.7 meter radio
telescope at 37\,GHz. The measurements were made with a 1\,GHz-band
dual beam receiver centered at 36.8\,GHz. The observations are ON-ON
observations, which alternate the source and the sky in each feed horn. A
typical integration time to obtain one flux density data point is
between 1200 and 1400\,s. A detailed description of the observation
and analysis methods can be found in \cite{terasranta}. 
The detection limit (defined as S/N $\geq$ 4) of the telescope is on the order
of 0.2\,Jy under optimal weather conditions. Given that the typical
flux density of \object{PKS 1424+240} at 37\,GHz is close to this limit, the
source can usually be observed during good weather only, and therefore,
the observed light curve from this source is  sparse. Figure~\ref{fig:lc}
(first panel from the top) shows the significant detections of the
source from 2009 to 2011. The measured flux densities are between
0.2-0.35\,Jy, which is close to the detection limit 
and results in large measurement errors.
Within these error bars, the light curve
does not show significant variability.

In addition to single dish observations, the source has also been
observed with the Very Long Baseline Array (VLBA) at 15\,GHz as a part of the $\gamma$-ray selected
MOJAVE sample \citep{lister}.
Analysis of all seven observed MOJAVE epochs up to now  reveals two moving components with
speeds of  55\,$\pm$\,24\,$\mu$arcsec/y and 51\,$\pm$\,9\,$\mu$arcsec/y (M. Lister, priv. comm.). 
Using  z =\,0.6, this converts to a speed of $\sim1.2\,c$ and Doppler factor $\delta=5.5$ 
(assuming an average BL Lac object viewing angle of $5^\circ$ as determined by  \cite{hovatta09}.  
Note that a viewing angle of $1^\circ$ results in $\delta=12$, which is rather slow but 
is in accordance with high-energy peaking BL Lac objects
showing lower Doppler factors in the $\gamma$-ray loud AGN \citep{lister}.
The VLBA core is resolved in the MOJAVE images and has a mean-fitted FWHM 
Gaussian of 0.1 milliarcsec, which corresponds to 0.67\,pc at z=0.6.
The images were also investigated for limb brightening,
which is considered to be a signature of a spine-sheath structure, and 
recent observations of Mrk 501, for example, have shown limb brightening \citep{piner}. 
However, for PKS~1424+240 we find no signature of limb brightening.

Simultaneous single-dish and VLBA observations show that the extended jet
contributes $\sim$60\,mJy to the total single dish 15 GHz flux density (M. Lister, priv. comm.).

\subsection{Optical data}
The source has been observed as part of the Tuorla blazar monitoring
program since 2006. The observations are done with the KVA (Kungliga Vetenskapsakademi) telescope
on La Palma, which is operated remotely from Finland and the Tuorla 1\,meter telescope, located in Finland. KVA consists of
two telescopes, the larger one being a 60\,cm (f/15) Cassegrain
telescope equipped with a CCD polarimeter capable of polarimetric
measurements in BVRI-bands that uses a plane-parallel calcite plate and a
super-achromatic $\lambda$/2 retarder. The second, a 
35\,cm Celestron telescope, can be
used for photometric measurements in B, V, and R-bands. The
observations of \object{PKS 1424+240} are done in the R-band and analyzed using the standard procedures with the pipeline developed for the monitoring program (Nilsson et al. in prep.).
The magnitudes are measured with differential photometry by comparing star magnitudes
from \cite{fiorucci}. 
The polarimetric measurements were done
without filters to improve the signal-to-noise of the observations. The degree of polarization
and position angle were calculated from the intensity
ratios of the ordinary and extraordinary beams using standard
formulae and semiautomatic software specially developed for
polarization monitoring purposes.

In April 2011 during the high optical state of the source, observations were also performed at the Nordic Optical Telescope to get a better sampling of the light curve, as KVA was suffering from technical problems. The data were reduced using the same procedure as for KVA and Tuorla data.

The object \object{PKS 1424+240} has also been observed as part of the Catalina Sky Survey (CSS) Program in the optical V-band. The data  are publicly available \citep{drake} \footnote{http://nesssi.cacr.caltech.edu/DataRelease/} and  are used to get better sampling for the optical light curve. 

During the six years of monitoring, the optical R-band magnitude of the
source was between 13.6 and 14.5 (V-band: 13.9 and 14.6), and
after the beginning of 2009, the source has been brighter than R=14.2 in
all measurements (V-band: 14.3). Historical data from 1994--1995
\citep{fiorucci} and  1988 \citep{mead90}
show R-band magnitudes from 14.75 (1988) to 14.2 (1995) and the V-band measurement from 1984 shows V=16.2. The optical
magnitudes measured from 2009 to 2011, therefore, clearly
present an optical high state of the source when compared
to historical data. Within these three years of data, the
average magnitude has been R$\sim14$ (7.7\,mJy) with 
the highest optical flux in 2011. The maximum
flux reached R=13.65 (10.7\,mJy), which is $\sim$40\% above
the average core flux from 2009-2011, and is the highest optical flux measured from the source to our knowledge.
 The 2011 MAGIC observations were triggered by this high optical flux. However, the  data taking of MAGIC started when the optical flux was already decreasing and the
average optical flux during 2011 MAGIC observations
was R=13.83 (9.0\,mJy), while  it was
R=14.01 (7.68\,mJy) and R=14.05 (7.39\,mJy) in 2009 and 2010, respectively. 

The optical light curves in the R- and V-bands start three years earlier
than the 15\,GHz light curves, but the increasing trend seems to be
present also in optical data. We also test for the existence of this trend 
in radio data. In the optical R-band, the trend is significant at
the $19\sigma$ level. The intrinsic modulation index of the light curve is
$0.152^{+0.010}_{-0.012}$ and with the trend subtracted it is
$0.071\pm0.005$. The same analysis is repeated for the V-band data
(which have better sampling in 2006-2008), resulting in a lower but still
highly significant trend at the $9\sigma$ level.

In 2011, we also performed two polarization measurements: May 6th (MJD
55688) and May 21st (MJD 55703). The polarization was found to be $7.2\pm 0.5$\% and $9.1\pm 0.6$\%, respectively. 
Compared to historical observations
from 1984 and 1988 (without filter $4.7\pm0.3$\% \citep{impey90}
and R-band $4.9\pm0.3$\% \citep{mead90}), 
the polarization is significantly
higher. The polarization degree is still within the typical range for BL Lacs
\citep[e.g.][]{jannuzi}.

\subsection{{\it Swift} UV and X-ray observations}

{\it Swift} observations of \object{PKS 1424+240} were performed using two of the three on-board instruments: The X-ray telescope 
\citep[XRT,][]{burrows05} 
covers the 0.2 - 10\,keV energy band, and the UV/Optical Telescope \citep[UVOT,][]{roming}
covers the 180 - 600 nm wavelength range with V, B, U, UVW1, UVM2, and UVW2 filters. The third instrument, the Burst Alert Telescope \citep[BAT,][]{barthelmy05}, is a coded-mask imager that covers the 15 - 150\, keV energy range. The source has not been detected by the complete analysis of 54 months of the BAT survey data, or the 2nd Palermo BAT catalog \citep{cusumano}, and therefore, the BAT data were not used in the present analysis. In the following, the XRT data from 0.2 to 10 keV  are discussed  with the UVOT data. 

{\it Swift} observed the source extensively in June 2009, following the detection of VHE emission by VERITAS \citep{ver_atel,acciari10}; a few sporadic observations followed in November 2009, January 2010, and November 2010, summing up to a total of nearly 26~ks of observations. Table~\ref{tab:SwiftLog} summarizes the {\it Swift} observations of \object{PKS 1424+240}.

The XRT data were processed using the FTOOLS task XRTPIPELINE (version 0.12.6), which is distributed by HEASARC within the HEASoft package (v6.10). Events with grades $0-12$ were selected for the data (see Burrows et al., 2005) and response matrices version 20100802 available in the {\it Swift} CALDB were used. For the spectral analysis, the modest pile-up affecting the June 2009 data was evaluated following the standard procedure\footnote{http://www.swift.ac.uk/analysis/xrt/pileup.php}, which resulted in a piled-up region with a radius of $\sim 7$ arcsec. This region was masked, and the signal was extracted  within an annulus with inner radius of 3 pixels (7.1 arcsec) and outer radius of  25 pixels (59 arcsec). 
The pile-up correction was applied only to observations 
with count rates higher than 0.6 cps.
The observations with ID 00038104001 and 00038104010 show a strip of dead pixels crossing the source; although the xrtpipeline creates an ancillary response file (ARF) correcting for the exposure map, we excluded these data from our analysis. The observations with ID 00039182002 and 00039182003 have few counts due to the low source flux and short observation time. The low statistics do not allow us to perform a reliable spectral fit, and for this reason, they have been excluded.

The spectra were extracted from the corresponding event files and binned using GRPPHA to ensure a minimum of 27 counts per bin in a manner, so that the $\chi ^2$ statistic could  be used reliably. Spectral analyses were performed using XSPEC version 12.6.0. 

The results are summarized in Appendix A. Table~\ref{tab:SwiftXRT} shows the results of a power-law model fit to the data. Photo-electric absorption with a hydrogen column density fixed to the Galactic value $n_H = 3.1 \times 10^{20}\,$ cm$^{-2}$ \citep{kaberla05} was included in the model. 

\begin{figure*}[t!]
\begin{center}
\includegraphics[width=0.95\textwidth]{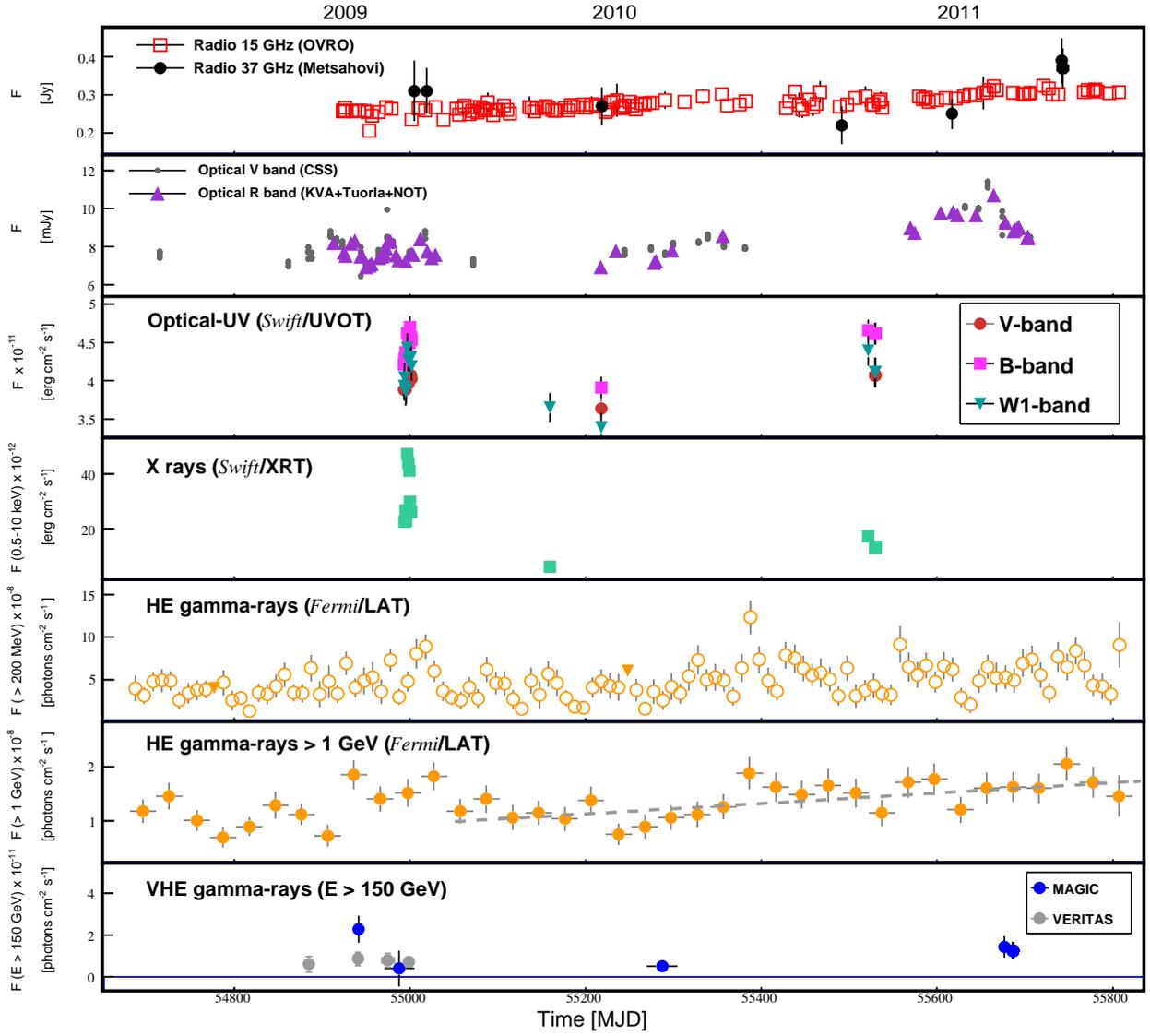}
\caption{Multiwavelength light curve of PKS~1424+240 from MJD 54700 to MJD 55800. Since all {\it Swift}/UVOT filter data show the same trend, only filters B, V, and W1  are shown. The dashed gray line represents a linear fit to the {\it Fermi}-LAT data after the X-ray flare. \label{fig:lc}}
\end{center}
\vspace{0.5cm}
\end{figure*}

At the beginning of the observations, the source was in a high X-ray state compared to the lower X-ray states in 2010. A flare has been observed on June, 15 2009 lasting 3 days (Figure~\ref{fig:lc}) with the flux doubling on a one-day timescale. During the burst, a spectral hardening is recognizable, although a clear harder-when-brighter trend cannot be seen in the whole data set.

The observed day-scale X-ray variability sets an upper limit for the size of the X-ray emission region given by the causality relation: $R~<~c t_{var}\delta /(1+z) \sim 1.6 \times 10^{15}\delta$\,[cm] by assuming a redshift of 0.6.


{\it Swift}/UVOT observed the source with all filters (V, B, U, UVW1, UVM2, UVW2) over 12 days. During three of those days, the filter-of-the-day mode was used. The UVOT source counts were extracted from a 5 arcsec-sized circular region centered on the source position, while the background was extracted from a nearby larger, source-free, circular region.
This data were processed with the {\it uvotmaghist} task of the HEASOFT package. The observed magnitudes have been corrected for Galactic extinction $E(B-V) = 0.059 \,$mag (Schlegel et al., 1998), applying the formulas by Pei (1992) and finally converting them into fluxes following Poole et al. (2008).
The observed magnitudes and the de-reddened fluxes are collected in Table A.3. The UV brightness is marginally variable, increasing during the X-ray flare of June 15.

\begin{figure*}[t!]
\begin{center}
\includegraphics[width=0.95\textwidth]{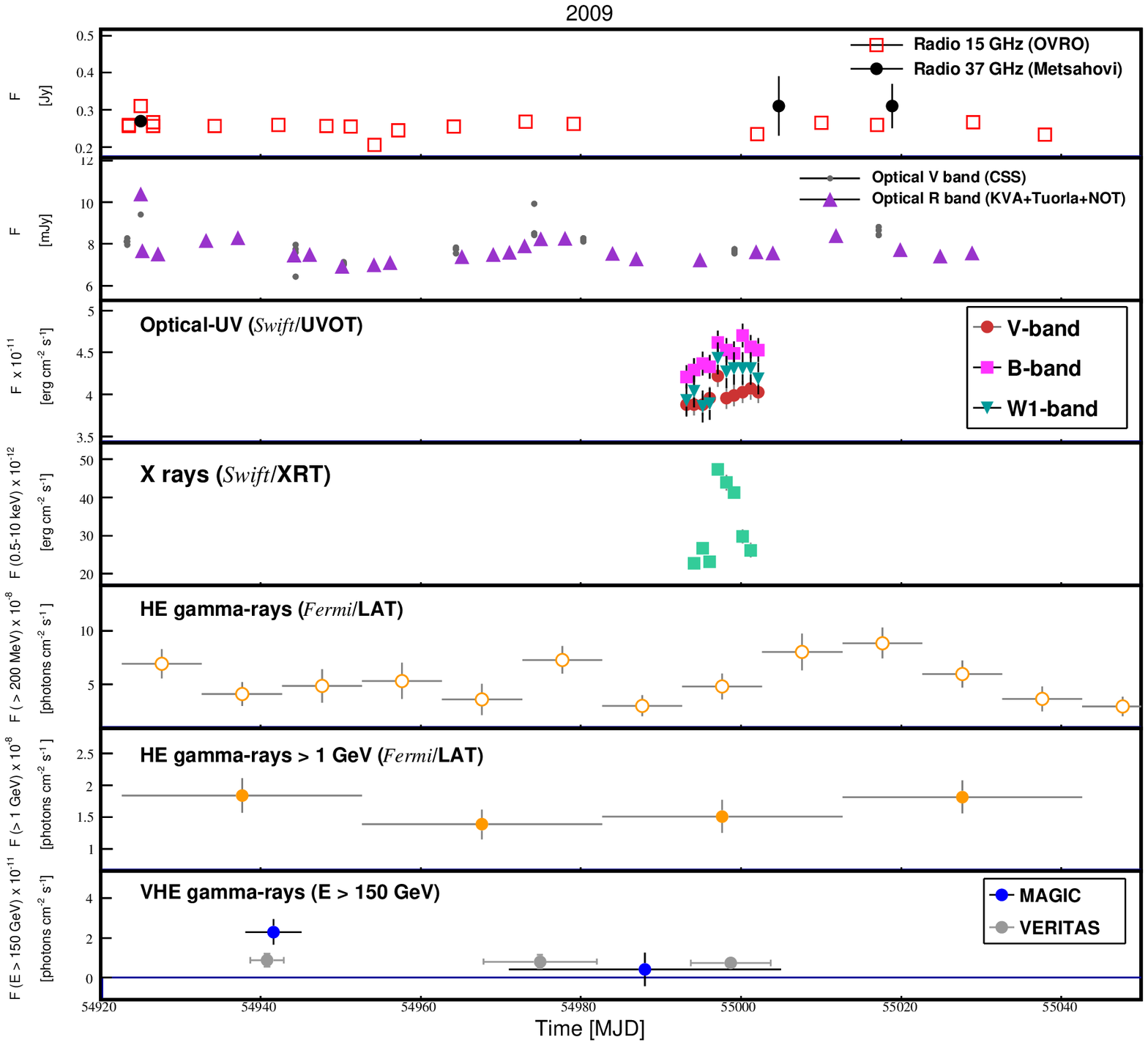} 
\caption{Zoom of multiwavelength light curve of PKS~1424+240 around the X-ray flare in 2009. \label{fig:lc_zoom}}
\end{center}
\vspace{0.5cm}
\end{figure*}

\subsection{{\it Fermi}-LAT $\gamma$-ray data}

{\it Fermi}-LAT is a pair-conversion $\gamma$-ray telescope sensitive to photon energies greater than 20\,MeV. In its nominal scanning mode, it surveys the whole sky every 3 hr with a field of view of about 2.4 steradians \citep{atwood09}.
The LAT data presented in this paper  were collected from MJD 54682 (2008 August 4) to MJD 55200 (2011 June 4).
An unbinned likelihood spectral analysis was performed to produce the light curve with the standard analysis tool {\it gtlike}, which included in the Science Tools software package (version v09r23p01)\footnote{For a documentation of the Science Tools, see http://fermi.gsfc.nasa.gov/ssc/data/analysis/documentation/ .}.
The P7\_SOURCE\_V6 instrument response functions  were used, which is a refinement to previous LAT response functions, reflecting improved understanding of the point-spread function and effective area \citep{ackermann12}. For this analysis, only photons located in a circular region of interest (ROI) with a 10 degree radius, centered at the position of PKS1424+240, were selected.
In addition, we excluded photons arriving from zenith angles $>$ 100\,$\degr$  to limit contamination from Earth limb $\gamma$ rays, and events are detected while the spacecraft rocking  angle was  $>$ 52\,$\degr$ to avoid time intervals during which Earth entered the LAT field of view (FoV).  

For the $\gamma$-ray analysis, we used the Galactic and Isotropic diffuse emission models  gal\_2yearp7v6\_v0.fits and iso\_p7v6source.txt.  The isotropic background is defined as  the sum of residual cosmic-ray background and extragalactic diffuse $\gamma$-ray background (file provided with Science Tools) \footnote{For details on the background models, see http://fermi.gsfc.nasa.gov/ssc/data/access/lat/BackgroundModels.html .}.
All point sources in the {\it Fermi}-LAT Second Source Catalog \citep[2FGL; ][]{nolan12} are within 20\,$\degr$ of PKS1424+240, including the source of interest itself, were considered in the analysis. Those within the  ROI were fitted with  power-law models with spectral indices set to the values obtained from the likelihood analysis of the full data set, while those beyond 10\,$\degr$ radius ROI had their values frozen to those found in the 2FGL.
Upper limits at 2-sigma confidence level were computed for time bins with test statistics (TS) \footnote{The test statistic \citep{mattox96} is defined as $TS = 2 (\log L_1 - \log L_0)$, where $L$ is the likelihood of the data given the model with ($L_1$) or without ($L_0$) a point source in the chosen position.}$<$ 9 and when the nominal flux uncertainty is larger than half the flux itself \citep{abdo10}. 
 All the uncertainties  stated here are statistics only,  the estimated systematic uncertainty of the integral fluxes above 200 MeV and 1 GeV are both about 10\% for a hard source like \object{PKS 1424+240} \citep{ackermann12}.


The spectral analysis was performed in the full band (from 100 MeV to 300 GeV)  for the whole 2008-2011 period using a simple power-law model. The best-fit parameters for the full-band model are $\Gamma = 1.784 \pm 0.016$ and an integral flux of $(7.7\pm 0.2)\times10^{-8}$ ph~cm$^{-2 }$s$^{-1}$ with the corresponding detection significance given by the $\sqrt{TS}\sim 86\sigma$.

The \emph{Fermi}-LAT integral flux light curves  above 200\,MeV and 1\,GeV 
were derived by performing separate flux estimations with 10- and 30-day bins, respectively, and 
are shown together with the multiwavelength light curves.
 They show a flickering behavior, and no strong activity is detected during the
investigated period. A constant flux hypothesis has a
low probability ($\chi^2/d.of.=225/110$, 
incompatible with the hypothesis of constant flux
at 95\% confidence level). 
The light curves have a hint of a fast flare occurring at the
same time as the X-ray flare in 2009 that was detected, and another similar event
was observed in July 2010 (MJD~55748), but no X-ray or VHE observations are
available during this flare.
No spectral variation is observed,  even during the 
X ray flares. 
The fitted value of the spectral index is consistent with being constant in time with a probability value of 93\%, 
which agrees with the hypothesis that there is  no significant 
spectral variations at a 95\% confidence level.

\subsection{Multiwavelength light curve}
The radio light curve at 15\,GHz and the optical light curves (R-
and V-bands) show a clear increase of flux with time, as concluded in Sections 
3.1 and 3.2. In the HE $\gamma$-ray light curves, the flux level 
seems to slightly increase toward the 2011 season.
In particular, for {\it Fermi}-LAT data above 1 GeV (after June 2009), 
the likelihood ratio test indicates a non-zero slope with more than 99\% confidence (dashed line in Fig.\ref{fig:lc}).
In X-rays and VHE $\gamma$ rays, the 
sampling is sparse and prevents measurement of any trend.
The trend seen in radio and optical  suggests a common
large, emission region for these wavebands. As it is known that most
of the emission at 15\,GHz originates in the parsec scale jet  (see Section 3.1), this also plays a significant part of the optical emission there.  However, in addition to the long-term trend, it is evident that the optical light curve also shows faster flares.

Figure~\ref{fig:lc_zoom} shows the light curves in 2009.
 There is fast X-ray flare observed in June 2009 and around this time, the flux in UVOT band also starts to increase. However, the increase is much slower and also smaller in amplitude, and therefore, it is not clear if it has common origin with the X-ray flare.

In the radio bands, the time coverage of the measurements does not allow any conclusion, and
in VHE and HE $\gamma$ rays, the simultaneous time bins do not show short-term variability. 
 Interestingly, there also seems to be a simultaneous increase of optical 
and HE $\gamma$-ray flux above  200 MeV at around the time when there was an X-ray outburst (MJD 54995), which might suggest a common origin for the optical and HE $\gamma$-ray emission. However, this trend is not significant in HE $\gamma$ rays when considering the entire data sample. 

\begin{figure}[htb]
\begin{center}
\includegraphics[width=0.45\textwidth]{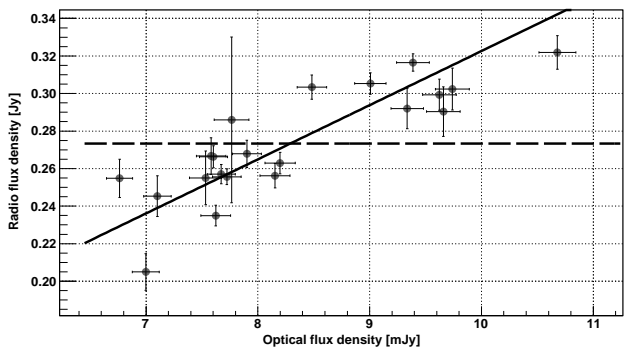}
\caption{Optical R-band flux versus radio 15\,GHz flux. The data pairs are selected with a time difference of $<0.9$ days. The linear fit is significantly preferred over the constant fit (see text for details). \label{fig:optradio_corr}}
\end{center}
\vspace{0.5cm}
\end{figure}

In addition to visual inspection of the light curves, we also perform
a simple correlation study between different bands. The only two bands
showing a possible correlation are the optical and radio (15\,GHz)
wavelengths: in Figure~\ref{fig:optradio_corr}, the data pairs of the
R-band and 15\,GHz with a time difference $< 0.9$ days are plotted
against each other. 
Although the linear fit has a low probability, it has a Spearman correlation coefficient of 81\% and, in a likelihood ratio test, it is significantly ($>99.9$\%)
 preferred over constant fit. 
The correlation is dominated by the long-term trend. We did not see any significant correlation when considering timescales shorter than one year. This result suggests that the radio and optical emission 
have a component in common that originates in a source, 
which varies on $\sim$1-year timescales. A confirmation of this statement will not be possible until radio/optical data are accumulated over many more years.

\subsection{Spectral energy distribution}
Using the multiwavelength data discussed in the previous subsections,
the SED of \object{PKS 1424+240} was constructed from
VHE $\gamma$ rays to the radio band. 
For VHE $\gamma$ rays, we plot the data from the
2009, 2010, and 2011 observations, which is corrected for the EBL absorption using the 
\cite{franceschini08} model. 
The EBL model of \cite{dominguez} gives compatible results.
The spectrum from {\it  Fermi}-LAT covers the whole 2008-2011 period. In X-rays,
we use two spectra from {\it Swift} from 2009, one from a high state (MJD
54997) and one from a low state (MJD 54996), and simultaneous UVOT data
for these epochs. The R-band points correspond to the average flux in
2009, 2010, and 2011 during the MAGIC observations. In the radio band,
we subtract  the contribution from the
extended jet  (see 3.1.) from the OVRO 15\,GHz measurement and plot the average flux between 2009 and
2011. Additionally, we use the {\it Planck} data from January-February
2010 that are published in \cite{giommi}. The data are not simultaneous, but
together they present a low and high state in different wavebands.

The SED of \object{PKS 1424+240} shows a wide
synchrotron bump peaking around the optical regime{\footnote {In the SED shown in Fig. 4 of \cite{acciari10}, the synchrotron peak seems to be narrower due to lower (possibly erroneously reproduced) X-ray luminosity.}.
 The location of the second peak is more uncertain  
but seems to be located at very high energies, making
the separation between the first and second peak
large. This feature is difficult to
model with a traditional single zone SSC. Additionally, the
multiwavelength variability suggests a possible correlation between
radio and optical fluxes, which means that the emission could originate
in the VLBA core for which the Doppler factor and the size
of the emitting region were derived from the MOJAVE data (see 3.1).
Therefore, in addition to a
traditional one-zone model \citep{tavecchio98, maraschi03} (Fig.~\ref{fig:sed}), we also
explore the two-zone model of \cite{tavecchio11} (Fig.~\ref{fig:sed2}). 
The parameters of the fits are presented in Table~\ref{tab:sed_param}.

In the canonical one-zone SSC model, the emitting plasma is contained in a
spherical region of radius $R$ in relativistic motion (described by a bulk
Lorentz factor $\Gamma$) along the jet at an angle $\theta$ with respect to the
line of sight to the observer, so that special relativistic effects
are cumulatively described by the relativistic Doppler factor, 
$\delta=[\Gamma(1-\beta cos \theta)]^{-1}$. The emitting region is filled with a homogeneous
tangled magnetic field with intensity $B$ and by a population of
relativistic electrons of density $n_e$, 
whose spectrum is described by a broken power law as a function of the energy of the relativistic electrons: 
\begin{equation}
N (\gamma)=K \gamma^{-n_1} \left(1+\frac{\gamma}{\gamma_b}\right)^{n_1-n_2},
\end{equation}
where $K$ is the normalization factor, $\gamma_b$ is the Lorentz factor of electrons at the spectral break and $n_1$ and $n_2$ the spectral indices below and above the break, respectively.

As a first approach, we fit the SED excluding the radio data 
(assuming that the radio emission originates in a different region 
 that does not contribute to the emission in other energy regimes). 
The fit is performed using the fully automatized
$\chi^2$-minimization procedure of \cite{mankuzhiyil11}. 
The systematic errors on the flux were estimated to be
2\%, 10\%, and 40\% for radio-optical-X-ray, GeV $\gamma$ rays, and VHE
$\gamma$ rays, respectively. 
The resulting fit is shown with a dashed line in Fig.~\ref{fig:sed}. 
While reproducing the high-energy SED data reasonably well, it fails 
in reproducing the shape of the optical-UV continuum (inset of Fig.~\ref{fig:sed}).
If the radio-optical connection can be confirmed with long-term study,
this model is disfavored because the radio emission is assumed to be produced in a different part of the jet. 

As a second approach, we include radio data in the modeling by considering
 the possible connection we found between radio and
optical wavebands.

However, it is not possible to reproduce the SED using a
Doppler factor of $\sim$10, such as that derived from the VLBA data, by assuming typical viewing angles of $1-5^{\circ}$ (see section 3.1).
The Doppler factor found from fitting the SED is an order of magnitude 
higher than the VLBA-derived value (dashed-dotted line fit of Fig.~\ref{fig:sed}). 
Even when constraining the Doppler factor to 40 in the fitting procedure, the low energy peak data cannot be properly described by the model (continuous line of 
Fig.~\ref{fig:sed}).  
Moreover, this last model cannot fulfill all the requirements. 
In particular, the variability timescale of one day observed during the X-ray 
flare in 2009 cannot be attained. 
The causality relation $(R < ct_{var}\delta/(1+z))$
gives  $R<6.5\times10^{16}$\,cm in this case, while the fit parameter requires $R\sim19\times 10^{16}$\,cm.
It, therefore, seems that 
the emission zone parameters do not agree with the ones 
derived for the VLBA core in the one-zone modeling. 
This is problematic because all the emission in a one zone model  should 
originate in the same region, which, basing on the radio emission, could be
placed in the VLBA core.

For the two-zone modeling, we adopt the model presented for the
flat-spectrum radio quasar PKS~1222+216 in \cite{tavecchio11} with
small modifications. We neglect the external seed photon contributions
as PKS1424+240 is a BL Lac object where the disk and broad-line region are
assumed to be weak, and therefore, external Compton models are not
appealing. The two emission regions are separated: the inner region
is assumed to be smaller. This  could mean 
that it is closer to the
central black hole or that it is a  spine, while the larger region is a sheath.
For the larger emission
region, the radius $R$ is equal to half the cross-sectional size of a conical
jet. 
Both regions are described by the same parameters
as the one-zone model, assuming an electron population 
distributed as a broken power-law in each region.
Since the cooling time of the electrons at the highest energies is shorter than the light-crossing time (i.e. the escape from the source), the absence of a break in the electron distribution requires a continuous acceleration of high-energy electrons.
The SED is modeled with two emission regions such
that the larger emission region would have the properties observed for
the VLBA core. The resulting fit is shown in Figure~\ref{fig:sed2}, and the adopted parameters are in the last line of Table~\ref{tab:sed_param}. The two-zone model results in an acceptable fit to the data. 
The high-energy part of the SED model
Figure~\ref{fig:sed2_he} reveals that in this case the peak of the second bump is located  at $\sim$60\,GeV.
The high $\gamma_{min}$ obtained is compatible with the hypothesis of shock heating, 
as suggested, for example, in  \citet{virtanen} and \citet{sironi}.

The measured powers indicate a strange behavior of the magnetic field, 
whose relative importance compared to kinetic energy of particles increases with the distance. For a very different radius R,
 we obtain the same magnetic field B, meaning a very large magnetic energy in the large zone. 
The behavior of the bulk Lorentz factor $\Gamma$ suggests some deceleration of the flow 
\citep[e.g.][]{georganopoulos} which could explain the magnetic field increase (through compression). 
This behavior, however, is not sufficient to explain the values of the model. Assuming a steady jet
under expansion, B is expected to scale as R$^{-1}$. Then the relation between the magnetic field and the 
radius of the inner and outer regions should be $B_{out}/B_{in}=R_{in}/R_{out}$=0.025. 
Deceleration could in principle compensate for this decrease, since  $B\propto \Gamma^{-1}$ through compression
if the decelerating jet propagates cylindrically. In the case of \object{PKS 1424+240}, if we consider that the viewing angle 
$\theta_{in}$=1/$\Gamma_{in}$, then $\Gamma_{in}=\gamma_{in}$, while the viewing angle of the outer region would 
be $\theta_{out}=1.91^\circ$ and therefore, $\Gamma_{out}=4.66$. The relation among the two emitting regions 
would be $B_{out}/B_{in} = \Gamma_{in}/\Gamma_{out} = 6.44$, which is too small (by a factor of ~10) to compensate for the expansion.
A possible explanation is a jet with primary jet injection that changes parameters with time. 
Alternatively, if the two regions are not located at different distances along the jet 
but are cospatial and form a spine-layer structure \citep[e.g.][]{ghisellini05},
it is then possible that the magnetic field is uniform in the jet, which 
 explains the same value in the fast and slow region.

\begin{table*}[th]
\small
\centering
\begin{tabular}{lcccccccccccc}
\hline
\hline
Model&$\gamma _{\rm min}$ & $\gamma _{\rm b}$ & $\gamma _{\rm max}$ &
$n_1$ & $n_2$ &$B$ & $K$ &$R$ & $\delta $ & 
L$_{kin(p)}$& L$_{kin(e)}$& L$_B$\\
&$[10^3$] & [$ 10^4$] &[$ 10^5$]  &  & &[G] & [cm$^{-3}]$  & $[10^{16}$cm] 
&&[$10^{45}$ erg s$^{-1}$]&[$10^{45}$ erg s$^{-1}$]&[$10^{43}$ erg s$^{-1}$]\\
\hline
one-zone  & 0.260 & 3.2 & 8.9$\times10^3$ & 1.9& 3.9 & 0.018 & 2$\times10^2$ & 6.5 & 70 & 5 & 7.0 & 3 \\ 
(no radio) &&&&&&&&&&&& \\
\hline
one-zone  & 0.016 & 2.6 & 3.9$\times10^2$ & 1.7 & 3.7 & 0.006 & 50 & 5 & 131 & 64 & 21 & 0.8 \\ 
\hline
one-zone &0.004&5.3 & 3.2$\times10^4$ &2.0 &4.0& 0.017&1.7$\times10^2$&19&40& 371 & 11 & 8.8 \\
(constrained) &&&&&&&&&&&& \\
\hline
\hline
2 zones  (in) & 8.0   &3.9& 7.0  &2.0&3.1&0.033&3.1$\times10^3$&4.8&30&0.07&1.2& 1.1\\
2 zones  (out)& 0.6 & 3.0  & 0.5&2.0&3.0&0.033 & 23 & 190 &9 & 1.3 & 2.3 & 159\\ 
\hline
\hline
\end{tabular}
\vskip 0.4 true cm
\caption{Model parameters for the four models: one-zone model without considering radio data, one-zone model including the radio data, one-zone model including the radio data and considering constraints derived from the VLBA data, and two-zone model where the outer emission region considers the constraints derived from the VLBA data.  
The following quantities are reported: the minimum, break, and maximum
Lorentz factors ($\gamma _{\rm min}$, $\gamma _{\rm b}$, $\gamma _{\rm max}$); the low and high-energy slope of the electron
energy distribution ($n_1$ and $n_2$);
the magnetic field intensity; $B$, the electron density; $K$, the radius of the
emitting region, $R$;  the Doppler factor, $\delta $; and the kinetic energy of the protons, electrons, and magnetic field (L$_{kin(p)}$, L$_{kin(e)}$, and L$_B$).}
\label{tab:sed_param}
\end{table*}

\begin{figure}
\begin{center}
\includegraphics[width=0.45\textwidth]{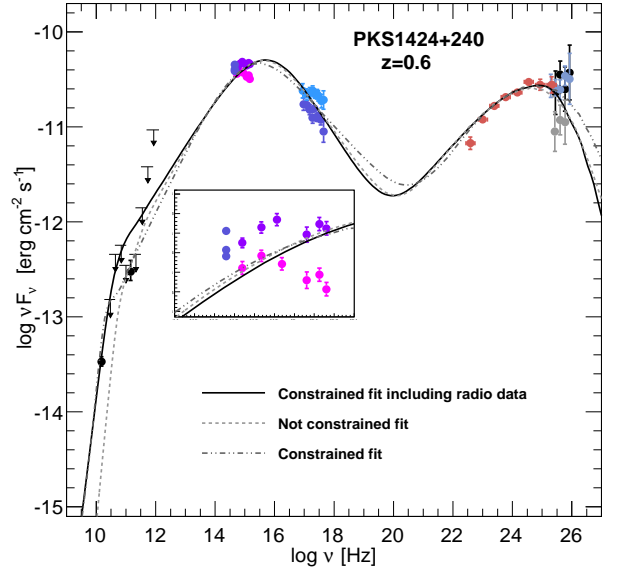}
\caption{SED of PKS~1424+240 constructed from radio to VHE $\gamma$ rays: radio 15\,GHz from OVRO (black filled circles), 30\,GHz to 857\,GHz from Planck (black arrows and filled circle), optical R-band from KVA (blue filled circles), optical to UV from UVOT (pink (lowest state) and purple (highest state) filled circles), X-rays from {\it Swift}-XRT (light blue (high state) and violet (low state) filled circles), HE $\gamma$ rays from {\it Fermi}-LAT (red filled circles) and VHE $\gamma$ rays from MAGIC (2009: black, 2010: gray, 2011: light blue).
The SED is fitted with three single-zone SSC models: the high $\gamma_{min}$ fit (solid line) and the fits that result from the $\chi^2$ minimization (dashed and dot-dashed lines) (see text for details).  The inset shows the optical-UV range of the SED on an expanded scale. A redshift of z~=~0.6 is assumed.}\label{fig:sed}
\end{center}
\vspace{0.5cm}
\end{figure}

\begin{figure}
\begin{center}
\includegraphics[width=0.45\textwidth]{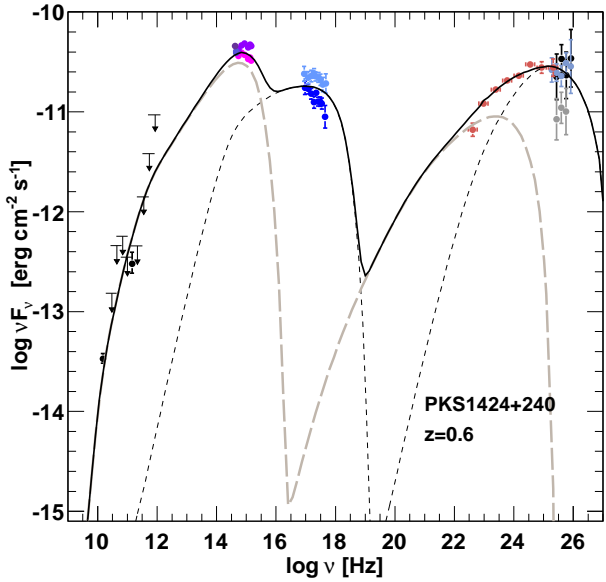}
\caption{SED of PKS~1424+240 fitted with a two-zone SSC model assuming a redshift of z~=~0.6. The long dashed line represents the emission from the outer region and the dashed line from the inner region (see text). Data as in Fig.~6. \label{fig:sed2}}
\end{center}
\vspace{0.5cm}
\end{figure}

\begin{figure}
\begin{center}
\includegraphics[width=0.45\textwidth]{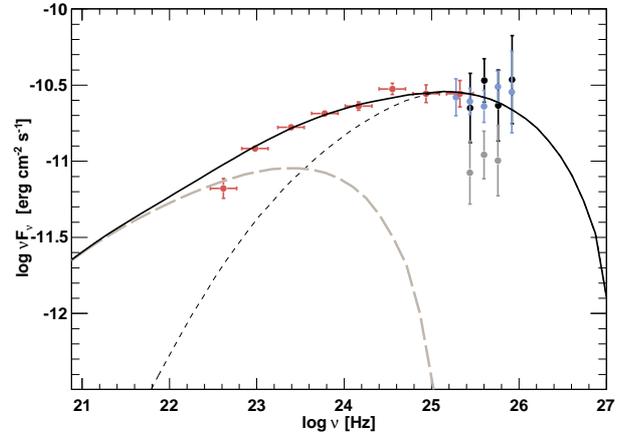}
\caption{High-energy part of the SED of PKS~1424+240 fitted with a two-zone SSC model and with {\it Fermi}-LAT data (red symbols) and MAGIC 2009, 2010, and 2011 data (gray, black, and blue symbols, respectively), which are corrected for EBL absorption by assuming redshift of 0.6 and the EBL model of \cite{franceschini08}. The peak is located at $\sim$60\,GeV. \label{fig:sed2_he}}
\end{center}
\vspace{0.5cm}
\end{figure}

\section{Summary and discussion}

 In this paper, we present the first long-term study of the VHE $\gamma$-ray
emission from the BL Lac object \object{PKS 1424+240}. The redshift of this source is uncertain.
Recently, a lower limit of the redshift z $\geq$ 0.6035 was determined from the Ly$\beta$ and Ly$\gamma$ absorption of the intergalactic medium by \cite{furniss}, making the source a strong candidate to be the farthest known VHE $\gamma$-ray emitter.

 MAGIC observed the source for three years from 2009 to 2011. We did not find any significant $\gamma$-ray excess above 400 GeV, while VERITAS did \citep[see][]{acciari10}, but this could be due to the shorter MAGIC observations (in comparison to those from VERITAS). The upper limit on the flux at 500 GeV reported in Figure 1 agrees with the value reported by \citet{acciari10} at the same energy. The 2 $\sigma$ confidence upper limit for the redshift, z $<$ 0.81, derived from the MAGIC 2011 spectrum (which is the most precise among the MAGIC spectra) and  the Franceschini et al. (2008) EBL model, agrees with the new lower limit.

\citet{furniss} reported that the VERITAS VHE spectrum (corrected for EBL absorption) exhibited a lower flux  than an extrapolation from the {\it Fermi}-LAT power-law spectrum in all but the highest energy bin, which was interpreted as an overestimation of the $\gamma$-ray opacity or the onset of an unexpected VHE spectral feature. 
In this work, we used the average 2008--2011 {\it Fermi}-LAT spectrum, which extends to higher energies and has smaller statistical error bars than the 2009 {\it Fermi}-LAT spectrum that is simultaneous to the VERITAS observations used in \citet{furniss}. The 3-year  Fermi-LAT spectrum extends beyond 100 GeV and connects smoothly (within uncertainties) with the 2009, 2010, and 2011 MAGIC spectra.


The VHE light curve of the source is compatible with the  
hypothesis of constant flux at a 95\% confidence level, and
the spectral index of the differential energy spectrum is steady within the
error bars. 
The MAGIC VHE $\gamma$-ray observations in 2009 and 2011 were
triggered by the optical outbursts in the source, but we find no
conclusive evidence for a higher VHE $\gamma$-ray flux from the source
during the optical outbursts in yearly timescales.
The two periods that were triggered by optical outburst showed
moderately higher VHE $\gamma$-ray flux than the 2010 period. However,  the
average optical flux was higher in 2011, while the highest VHE
$\gamma$-ray flux was measured in 2009. 
As the source is rather weak
in the VHE $\gamma$-ray band and the observations of several nights
have to be combined in the light curve, a detailed short term
comparison is not possible. 

We also present the first long-term multiwavelength dataset of the
source that covers  2006-2011 and includes data from radio to HE
$\gamma$ rays. The source is variable in all studied energy regimes,
although the variability is moderate. We find a
significant  increasing trend in the 15\,GHz radio  and optical
light curves in 2009-2011, as well as correlation between the two
bands, which we interpret as a common large, emission region,
which possibly could be the parsec scale jet. We investigate the
MOJAVE 15\,GHz VLBA observations of the source and find that the
parsec scale jet is  slow and therefore could be responsible for
such long-term variability. The common origin of radio and optical
emission in blazars has been studied extensively in the past
\cite[e.g.][]{tornikoski94,hanski02}. The results have been 
inconclusive with some sources showing correlations while others
not. Here, we find a clear indication of connection between these two
energy regions in a high synchrotron peaking source. Sources of this type are weak in
the radio band, and therefore, such a connection has not been considered in the past
studies. We emphasize the need for further investigation of this
connection in a sample of sources.

We also investigate the SED of the source
with a non-simultaneous dataset that presents the low and high state of
the source within 2009-2011. 
The one-zone fit to the SED (both including and excluding the radio data) fails to 
reproduce the optical-UV continuum and
requires  a much higher Doppler factor than that observed in the parsec scale jet by the
VLBA. Even when constraining the Doppler factor in the fitting procedure, 
no result is derived for a Doppler factor lower than 40.
The same mismatch in Doppler factors has been seen for many VHE
$\gamma$-ray emitting BL Lac objects \citep[see, e.g.,][and references
  therein]{tiet}. In many of the sources, the high Doppler factor of
the VHE $\gamma$-ray emitting region is  required by the
fast variability seen in the VHE $\gamma$-ray regime \citep{albert07,
  aharonian07}. This indicates separate emission regions for radio and
$\gamma$-ray emission, and it has been suggested that jets are
deccelerating \citep{georganopoulos} or have spine-sheath structure
\citep{ghisellini05}. 
In the case of \object{PKS 1424+240}, the one-day scale variability is only observed in X-rays, which limits the size of the emission region to $\sim 1.6\times 10^{15} \delta$\,[cm].
Given the multiple pieces of
evidence in favor of two emission zones, we model the SED
with the two-zone model. According to this model, the radio and  majority
 of the optical emission originate in a large emission region,
and the X-ray to $\gamma$-ray emission originates mainly
in a smaller emission region with a minor contribution from the 
larger region in the {\it Fermi}-LAT energy band. 
The two-zone model has twice as many parameters 
and  fits  the SED better than the single zone models. However, this model requires a high $\gamma_{min}$
for the inner
region that produces the $\gamma$-ray emission 
 to not  overproduce the synchrotron emission in the optical and radio bands. 
As discussed in \cite{on325}, high $\gamma_{min}$ values have
also been adopted for other blazars, like 1ES~1215+303 and the so-called
extreme BL Lacs \citep[that show very hard X-ray spectra;][]{katarzynski, tavecchio09, kaufmann11, lefa11}, 
and are consistent with simulation results \citep{virtanen,sironi}.
Other evidence in favor of the two-zone model is that the X-ray flare is less visible in V and R-bands. The model can account for this feature, since a higher flux in the inner (small) component will mainly affect UV frequencies and higher frequencies in the optical, while the V and R-bands are more dominated by the emission from the larger component.

Based on the long-term multiwavelength light curve studies and the SED modeling, we conclude that the two emission region model is favored for \object{PKS 1424+240}, while the previous work \citep{acciari10} concluded that the one-zone SSC model is sufficient for describing the SED. This shows that inclusion of long-term multiwavelength light curves from radio to $\gamma$ rays gives significant input for the modeling of the emission.
However, the SED studies here were done with the non-simultaneous 
data, and we emphasize the
need for strictly simultaneous multiwavelength observations for this
source as a starting point for further studies.

\begin{acknowledgements}
We are grateful to the anonymous referee for the valuable
comments and suggestions.
We would like to thank the Instituto de Astrofisica de 
Canarias for the excellent working conditions at the 
Observatorio del Roque de los Muchachos in La Palma. 
The support of the German BMBF and MPG, the Italian INFN 
and Spanish MICINN is gratefully acknowledged. 
This work was also supported by ETH Research Grant 
TH 34/043, by the Polish MNiSzW Grant N N203 390834, 
by the YIP of the Helmholtz Gemeinschaft, and by grant DO02-353
of the the Bulgarian National Science Fund.

Partly based on observations made with the Nordic Optical Telescope, operated
on the island of La Palma jointly by Denmark, Finland, Iceland,
Norway, and Sweden, in the Spanish Observatorio del Roque de los
Muchachos of the Instituto de Astrofisica de Canarias. 
The data presented here were obtained in part with ALFOSC, which is 
provided by the Instituto de Astrofisica de Andalucia (IAA) under a joint 
agreement with the University of Copenhagen and NOTSA.

The CSS survey is funded by the National Aeronautics and Space
Administration under Grant No. NNG05GF22G issued through the Science
Mission Directorate Near-Earth Objects Observations Program.  The CRTS
survey is supported by the U.S.~National Science Foundation under
grants AST-0909182. 

The OVRO 40-m monitoring program is
supported in part by NASA grants NNX08AW31G 
and NNX11A043G, and NSF grants AST-0808050 
and AST-1109911. 

The \textit{Fermi} LAT Collaboration acknowledges generous ongoing support
from a number of agencies and institutes that have supported both the
development and the operation of the LAT as well as scientific data analysis.
These include the National Aeronautics and Space Administration and the
Department of Energy in the United States, the Commissariat \`a l'Energie Atomique
and the Centre National de la Recherche Scientifique / Institut National de Physique Nucl\'eaire et de Physique des Particules in France, the Agenzia 
Spaziale Italiana and the Istituto Nazionale di Fisica Nucleare in Italy, 
the Ministry of Education, Culture, Sports, Science and Technology (MEXT), 
High Energy Accelerator Research Organization (KEK) and Japan Aerospace 
Exploration Agency (JAXA) in Japan, and the K.~A.~Wallenberg Foundation, 
the Swedish Research Council and the Swedish National Space Board in Sweden.
Additional support for science analysis during the operations phase 
is gratefully acknowledged from the Istituto Nazionale di Astrofisica in 
Italy and the Centre National d'\'Etudes Spatiales in France.
\end{acknowledgements}

\appendix
\section{Detailed {\it Swift}-UVOT and XRT Results}
\begin{table*}[htdp]
\caption{Summary of {\it Swift} observations on PKS 1424+240 in 2009-2010 with the two instruments, XRT and UVOT.}
\begin{center}
\begin{tabular}{ccc|ccc|cc}
\hline
Obs. ID & Date & start time & XRT expos. & XRT counts & Pile-up & UVOT exposure & Filters \\
   & & [MJD] & (PC-mode) [s] & [cnts s$^{-1}$] & correction & [s]& \\
 \hline
38104001 & 11-Jun-2009 & 54993.1840 & 1732 & 0.43 & y & 1707 & all \\
38104002 & 12-Jun-2009 & 54994.1181 & 1785 & 0.49 & y & 1736 & all\\
38104003 & 13-Jun-2009 & 54995.1965 & 2263 & 0.45 & y & 2215 & all\\
38104004 & 14-Jun-2009 & 54996.0674 & 2311 & 0.5 & y & 2261 & all\\
38104005 & 15-Jun-2009 & 54997.1333 & 2093 & 0.89 & y & 2044 & all\\
38104006 & 16-Jun-2009 & 54998.2153 & 1132 & 0.88 & y & 1106 & all\\
38104007 & 17-Jun-2009 & 54999.1570 & 1269 & 0.84 & y & 1220 & all\\
38104008 & 18-Jun-2009 & 55000.2132 & 1132 & 0.45 & y & 1110 & all\\
38104009 & 19-Jun-2009 & 55001.2208 & 1132 & 0.52 & y & 1108 & all\\
38104010 & 20-Jun-2009 & 55002.1695 & 1403 & 0.27 & n & 1356 & all\\
39182001 & 24-Nov-2009 & 55159.3083 & 4696 & 0.14 & n & 4691 & UW1\\
39182002 & 22-Jan-2010 & 55218.0097 & 1005 & 0.09 & n & 984 & all\\
39182003 & 22-Jan-2010 & 55218.0764 & 1476 & 0.09 & n & 1477 & UM2\\
40847001 & 21-Nov-2010 & 55521.7549 & 1946 & 0.56 & n & 1900 & B U W1 M2 W2\\
41539001 & 29-Nov-2010 & 55529.5202 & 1131 & 0.47 & n & 1105 & all\\
\hline
\end{tabular}
\end{center}
\label{tab:SwiftLog}

\end{table*}%

\begin{table*}[htdp]
\caption{Results of the analysis on {\it Swift -}XRT data.  The flux and spectral indices have been derived with a fit on XRT data in the range $0.5-10\,$keV using a simple power-law model with photo-electric absorption  ($WABS$ model in XSPEC  for $n_H = 3.1 \times 10^{20}\,$ cm$^{-2}$). }
\begin{center}
\begin{tabular}{ccc|cc|cc|c}
\hline
\hline
Obs. ID & Date & Exposure & Flux 0.5-10 keV & Flux err & index & idx err & Red. $\chi ^2$/NDF \\
 & [MJD]  & [s] & [10$^{12}$ erg cm$^{-2}$ s$^{-1}$] &  &    &  & \\
 \hline
\hline
38104001 &  54993.1840 & 1732 & -- & -- & -- & -- & --  \\
38104002 & 54994.1181 & 1785 & 22.7 & 1.1 & 2.54 & 0.09 & 1.23/25  \\
38104003 & 54995.1965 & 2263 & 26.7 & 1.1 & 2.55 & 0.09 & 1.00/28  \\
38104004 & 54996.0674 & 2311 & 23.1 & 1.1 & 2.55 & 0.09 & 1.33/30  \\
38104005 & 54997.1333 & 2093 & 47.3 & 1.5 & 2.37 & 0.06 & 1.03/52  \\
38104006 & 54998.2153 & 1132 & 43.9 & 2.1 & 2.41 & 0.10 & 1.08/28  \\
38104007 & 54999.1570 & 1269 & 41.3 & 1.0 & 2.42 & 0.09 & 1.15/30  \\
38104008 & 55000.2132 & 1132 & 29.8 & 1.8 & 2.55 & 0.14 & 1.58/13  \\
38104009 & 55001.2208 & 1132 & 26.2 & 1.9 & 2.34 & 0.16 & 1.7/15  \\
38104010 & 55002.1695 & 1403 & -- & -- & -- & -- & --  \\
39182001 & 55159.3083 & 4696 & 6.1 & 0.4 & 2.53 & 0.09 & 1.06/31  \\
39182002 & 55218.0097 & 1005  & -- & -- & -- & -- & --  \\ 
39182003 & 55218.0764 & 1476 & -- & -- & -- & -- & --  \\ 
40847001 & 55521.7549 & 1946 & 17.3  & 0.9 & 2.34 & 0.08 & 1.40/33\\
41539001 & 55529.5202 & 1131 & 13.2 & 1.1 & 2.36 & 0.14 & 1.23/15 \\
\hline 
\hline
\end{tabular}
\end{center}
\label{tab:SwiftXRT}
\end{table*}

\begin{sidewaystable*}[htdp]
\centering
\rule{0.\textheight}{0.745\textheight}
\begin{tabular}{c|cc|cc|cc|cc|cc|cc}
\hline
\hline
Obs. ID & \multicolumn{2}{c}{UVV} &  \multicolumn{2}{c}{UBB} &  \multicolumn{2}{c}{UUU} &   \multicolumn{2}{c}{UW1} &   \multicolumn{2}{c}{UM2} &   \multicolumn{2}{c}{UW2} \\
 Central $\lambda _nm$ & \multicolumn{2}{c}{547} & \multicolumn{2}{c}{439} & \multicolumn{2}{c}{346} & \multicolumn{2}{c}{260} &  \multicolumn{2}{c}{225} & \multicolumn{2}{c}{193} \\
\hline
\hline
 & Obs magn & Flux $\times 10^{-11}$& Obs magn & Flux $\times 10^{-11}$ & Obs magn & Flux $\times 10^{-11}$ & Obs magn & Flux$\times 10^{-11}$ & Obs magn & Flux $\times 10^{-11}$ & Obs magn & Flux $\times 10^{-11}$  \\
 & mag &  erg cm$^{-2}$ s$^{-1}$ & mag &  erg cm$^{-2}$ s$^{-1}$ & mag &  erg cm$^{-2}$ s$^{-1}$ & mag &  erg cm$^{-2}$ s$^{-1}$ & mag&  erg cm$^{-2}$ s$^{-1}$ & mag & erg cm$^{-2}$ s$^{-1}$ \\
\hline
\hline
38104001 & 14.48 & 3.88 & 14.81 & 4.21 & 13.94 & 4.24 & 13.94 & 3.93 & 13.89 & 4.21 & 13.98 & 4.10 \\
38104002 & 14.48 & 3.88 & 14.79 & 4.29 & 13.92 & 4.32 & 13.91 & 4.04 & 13.9 & 4.18 & 13.94 & 4.26\\
38104003 & 14.48 & 3.88 & 14.77 & 4.37 & 13.88 & 4.48 & 13.96 & 3.86 & 13.95 & 3.99 & 14.09 & 3.71\\
38104004 & 14.46 & 3.96 & 14.78 & 4.33 & 13.88 & 4.48 & 13.95 & 3.89 & 13.87 & 4.29 & 13.98 & 4.10\\
38104005 & 14.39 & 4.22 & 14.71 & 4.62 & 13.8 & 4.82 & 13.81 & 4.43 & 13.77 & 4.71 & 13.88 & 4.50\\
38104006 & 14.46 & 3.96 & 14.73 & 4.53 & 13.84 & 4.65 & 13.85 & 4.27 & 13.83 & 4.45 & 13.93 & 4.30\\
38104007 & 14.45 & 3.99 & 14.74 & 4.49 & 13.83 & 4.69 & 13.84 & 4.31 & 13.82 & 4.49 & 13.93 & 4.30\\
38104008 & 14.44& 4.03 & 14.69 & 4.70 & 13.81 & 4.78 & 13.84 & 4.31 & 13.8 & 4.58 & 13.92 & 4.34\\
38104009 & 14.43 & 4.07 & 14.72 & 4.57 & 13.8 & 4.82 & 13.84 & 4.31 & 13.8 & 4.58 & 13.92 & 4.34\\
38104010 & 14.44 & 4.03 & 14.73 & 4.53 & 13.88 & 4.48 & 13.87 & 4.19 & 13.85 & 4.37 & 13.94 & 4.26\\
39182001 & -- & -- & -- & -- & -- & -- & 14.02 & 3.65 & -- & -- & -- & --\\
39182002 & 14.55 & 3.64 & 14.89 & 3.91 & 14.08 & 3.73 & 14.1 & 3.39 & 14.09 & 3.50 & 14.25 & 3.20\\
39182003 & -- & -- & -- & -- & -- & -- & -- & -- & 14.14 & 3.35 & -- & --\\
40847001 & -- & -- & 14.7 & 4.66 & 13.79 & 4.87 & 13.82 & 4.39 & 13.82 & 4.49 & 13.93 & 4.30\\
41539001 & 14.43 & 4.07 & 14.71 & 4.62 & 13.87 & 4.52 & 13.89 & 4.11 & 13.91 & 4.14 & 14 & 4.03\\
\hline
\hline
Errors & 0.10 & 0.13 & 0.10 & 0.14 & 0.08 & 0.17 & 0.08 & 0.19 & 0.06 & 0.17 & 0.06 & 0.16\\
\hline
\hline
\end{tabular}
\caption{UV observed magnitudes and dereddened fluxes from {\it Swift -}UVOT data. The central wavelength correspondent to each filter is indicated. The flux has been corrected for the galactic absorption. Flux values and its errors are in $10^{-11}\,$erg/cm$^2$/s units. The uncertainties on the measurements are approximately constant for each filter and are thus shown in the last row. See text for details on data analysis and correction.}
\label{tab:SwiftUVOT}
\end{sidewaystable*}%
\end{document}